\documentclass[journal]{IEEEtran}
\usepackage{cite}
\usepackage{amsmath,amssymb,amsfonts}
\usepackage{algorithmic}
\usepackage{graphicx}
\usepackage{textcomp}
\usepackage{subfigure}
\usepackage{graphicx}
\usepackage{color}
\usepackage{makecell}
\usepackage{url}
\usepackage{threeparttable}
\usepackage{ulem}
\usepackage{lscape}
\usepackage{hyperref}
\usepackage{array}
\usepackage{booktabs}
\usepackage{colortbl}
\usepackage{multirow}
\def\BibTeX{{\rm B\kern-.05em{\sc i\kern-.025em b}\kern-.08em
    T\kern-.1667em\lower.7ex\hbox{E}\kern-.125emX}}

\usepackage{verbatim}
\usepackage{multirow}

\begin{document}

\title{Deep Learning on Traffic Prediction: Methods, Analysis and Future Directions}

\author{Xueyan Yin, Genze Wu, Jinze Wei, Yanming Shen, Heng Qi, and Baocai Yin

\thanks{Xueyan Yin, Genze Wu, Jinze Wei, and Heng Qi are with the School of Electronic Information and Electrical Engineering, Dalian University of Technology, Dalian 116024, China. }
\thanks{Yanming Shen is with the School of Electronic Information and Electrical Engineering, Dalian University of Technology, Dalian 116024, China, and also
with the Key Laboratory of Intelligent Control and Optimization for
Industrial Equipment, Ministry of Education, Dalian University of Technology,
Dalian 116024, China (e-mail: shen@dlut.edu.cn).}
\thanks{Baocai Yin is with the School of Electronic Information and Electrical
Engineering, Dalian University of Technology, Dalian 116024, China, and also
with the Peng Cheng Laboratory, Shenzhen 518055, China.} }

\IEEEtitleabstractindextext{%
\begin{abstract}
 Traffic prediction plays an essential role in intelligent transportation
 system. Accurate traffic prediction can assist route planing, guide
 vehicle dispatching, and mitigate traffic congestion. This problem is
 challenging due to the complicated and dynamic spatio-temporal
 dependencies between different regions in the road network. Recently, a
 significant amount of research efforts have been devoted to this area,
 especially deep learning method, greatly advancing traffic prediction
 abilities. The purpose of this paper is to provide a comprehensive survey
 on deep learning-based approaches in traffic prediction from multiple
 perspectives. Specifically, we first summarize the existing traffic
 prediction methods, and give a taxonomy. Second, we
list the state-of-the-art approaches in different traffic prediction
applications. Third, we comprehensively collect and organize widely used
public datasets in the existing literature to facilitate other researchers.
Furthermore, we give an evaluation and analysis by conducting extensive
experiments to compare the performance of different methods on a real-world
public dataset. Finally, we discuss open challenges in this field.

\end{abstract}

}

\maketitle
\IEEEdisplaynontitleabstractindextext
\IEEEpeerreviewmaketitle

\section{Introduction}
\IEEEPARstart {T}{he} modern city is gradually developing into a smart city.
The acceleration of urbanization and the rapid growth of urban population
bring great pressure to urban traffic management. Intelligent Transportation
System (ITS) is an indispensable part of smart city, and traffic prediction
is an important component of ITS. Accurate traffic prediction is essential to
many real-world applications. For example, traffic flow prediction can help
city alleviate congestion; car-hailing demand prediction can prompt
car-sharing companies pre-allocate cars to high demand regions. The growing
available traffic related datasets provide us potential new perspectives to
explore this problem.

\textbf{Challenges} Traffic prediction is very challenging, mainly affected
by the following complex factors{:}

\begin{figure}[ht]
\centering
\includegraphics[width=0.5\linewidth]{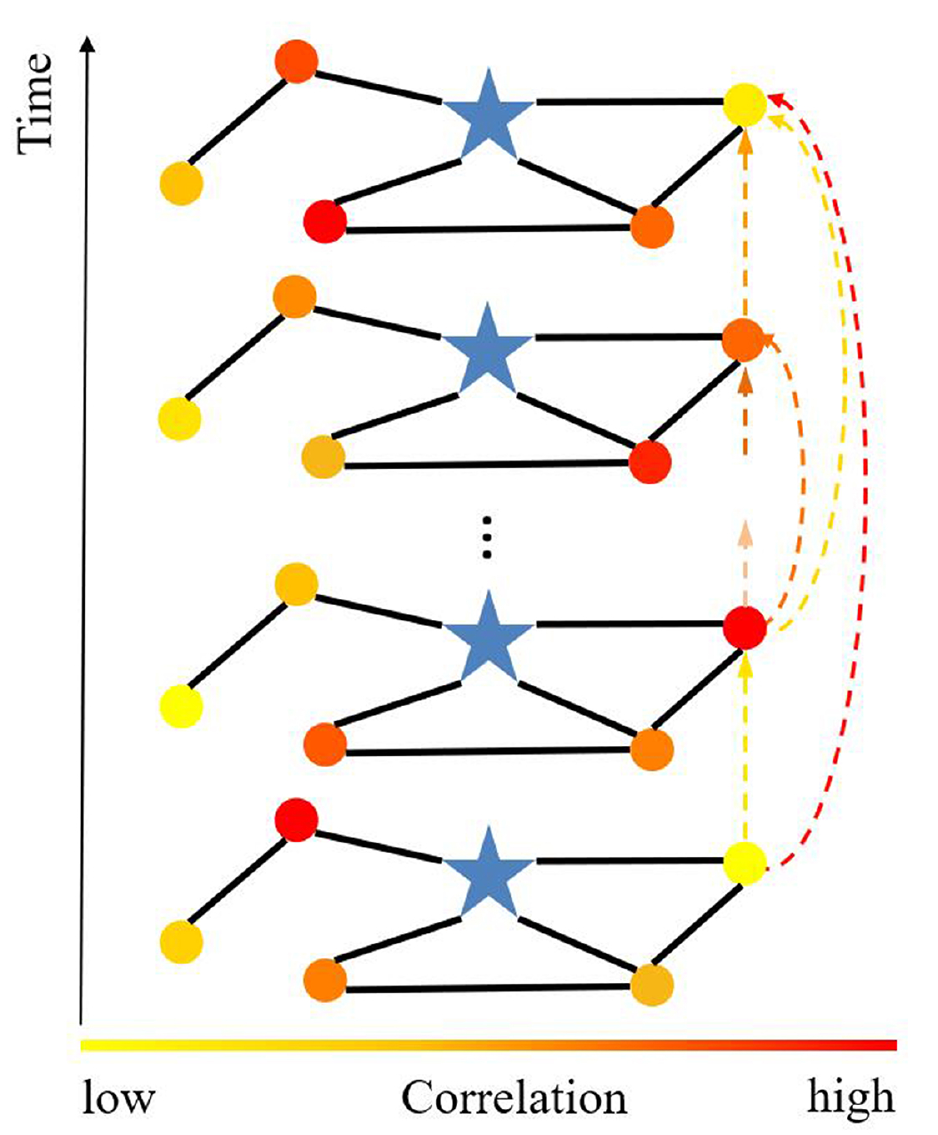}
\caption{Complex spatio-temporal correlations. The nodes represent different locations in the road network, and the blue star node represents the predicted target. The darker the color, the greater the spatial correlation with the target node. The dotted line shows the temporal correlation between different time steps.}
\label{challenge}
\end{figure}

(1) Because traffic data is spatio-temporal, it is constantly changing with
time and space, and has complex and dynamic spatio-temporal dependencies.
\begin{itemize}
\item Complex spatial dependencies. Fig.\ref{challenge} demonstrates that
    the influence of different positions on the predicted position is
    different, and the influence of the same position on the predicted
    position is also varying with time. The spatial correlation between
    different positions is highly dynamic.

\item Dynamic temporal dependencies. The observed values at different times
    of the same position show non-linear changes, and the traffic state of
    the far time step sometimes has greater influence on the predicted time
    step than that of the recent time step, as shown in
    Fig.\ref{challenge}. Meanwhile, \cite{zhang2016dnn} pointed out that
    traffic data usually presents periodicity, such as closeness, period
    and trend. Therefore, how to select the most relevant historical
    observations for prediction remains a challenging problem.
\end{itemize}
(2) External factors. Traffic spatio-temporal sequence data is also
influenced by some external factors, such as weather conditions, events or
road attributes.

Since traffic data shows strong dynamic correlation in both spatial and
temporal dimensions, it is an important research topic to mine the non-linear
and complicated spatial-temporal patterns, making accurate traffic
predictions. Traffic prediction involves various application tasks. Here, we
list the main application tasks of the existing traffic prediction work,
which are as follows:

\begin{itemize}
\item Flow

Traffic flow refers to the number of vehicles passing through a given point
on the roadway in a certain period of time.
\item Speed

The actual speed of vehicles is defined as the distance it travels per unit
of time. Most of the time, due to factors such as geographical location,
traffic conditions, driving time, environment and personal circumstances of
the driver, each vehicle on the roadway will have a speed that is somewhat
different from those around it.
\item Demand

The problem is how to use historical requesting data to predict the number
of requests for a region in a future time step, where the number of
start/pick-up or end/drop-off is used as a representation of the demand in
a region at a given time.
\item Travel time

In the case of obtaining the route of any two points in the road network,
estimating the travel time is required. In general, the travel time should
include the waiting time at the intersection.
\item Occupancy

The occupancy rate explains the extent to which vehicles occupy road space,
and is an important indicator to measure whether roads are fully utilized.



\end{itemize}

\textbf{Related surveys on traffic prediction} There are a few recent surveys
that have reviewed the literatures on traffic prediction in certain contexts
from different perspectives. \cite{Vlahogianni2014Short} reviewed the methods
and applications from 2004 to 2013, and discussed ten challenges that were
significant at the time. It is more focused on considering short-term traffic
prediction and the literatures involved are mainly based on the traditional
methods. Another work \cite{Li2018a} also paid attention to short-term
traffic prediction, which briefly introduced the techniques used in traffic
prediction and gave some research suggestions. \cite{nagy2018survey} provided
sources of traffic data acquisition, and mainly focused on traditional
machine learning methods. \cite{Singh2019Traffic} outlined the significance
and research directions of traffic prediction. \cite{boukerche2020machine}
and \cite{lana2018road} summarized relevant models based on classical methods
and some early deep learning methods. Alexander et al.
\cite{tedjopurnomo2020survey} presented a survey of deep neural network for
traffic prediction. It discussed three common deep neural architectures,
including convolutional neural network, recurrent neural network, and
feedforward neural network. However, some recent advancements, e.g.,
graph-based deep learning, were not covered in \cite{tedjopurnomo2020survey}.
\cite{ye2020build} is an overview of graph-based deep learning architecture,
with applications in the general traffic domain. \cite{xie2020urban} provided
a survey focusing specifically on the use of deep learning models for
analyzing traffic data. However, it only investigates the traffic flow
prediction. In general, different traffic prediction tasks have common
characteristics, and it is beneficial to consider them jointly. Therefore,
there is still a lack of broad and systematic survey on exploring traffic
prediction in general.

\textbf{Our contributions} To our knowledge, this is the first comprehensive
survey on deep learning-based works in traffic prediction from multiple
perspectives, including approaches, applications, datasets, experiments,
analysis and future directions. Specifically, the contributions of this
survey can be summarized as follows:
\begin{itemize}
\item We first do a taxonomy for existing approaches, describing their key
    design choices.
\item We collect and summarize available traffic prediction datasets, which
    provide a useful pointer for other researches.
\item We perform a comparative experimental study to evaluate different
    models, identifying the most effective component.
\item We further discuss possible limitations of current solutions, and
    list promising future research directions.
\end{itemize}

\textbf{A Taxonomy of Existing Approaches} After years of efforts, the
research on traffic prediction has achieved great progresses. In light of the
development process, these methods can be broadly divided into two
categories: classical methods and deep learning-based methods. Classical
methods include statistical methods and traditional machine learning methods.
The statistical method is to build a data-driven statistical model for
prediction. The most representative algorithms are Historical Average (HA),
Auto-Regressive Integrated Moving Average (ARIMA)\cite{williams2003modeling},
and Vector Auto-Regressive (VAR)\cite{zivot2006vector}. Nevertheless, these
methods require data to satisfy certain assumptions, and time-varying traffic
data is too complex to satisfy these assumptions. Moreover, these methods are
only applicable to relatively small datasets. Later, a number of traditional
machine learning methods, such as Support Vector Regression
(SVR)\cite{chen2015forecasting} and Random Forest Regression
(RFR)\cite{johansson2014regression}, were proposed for traffic prediction
problem. Such methods have the ability to process high-dimensional data and
capture complex non-linear relationships.

\begin{figure*}[ht]
\centering
\includegraphics[width=0.8\linewidth]{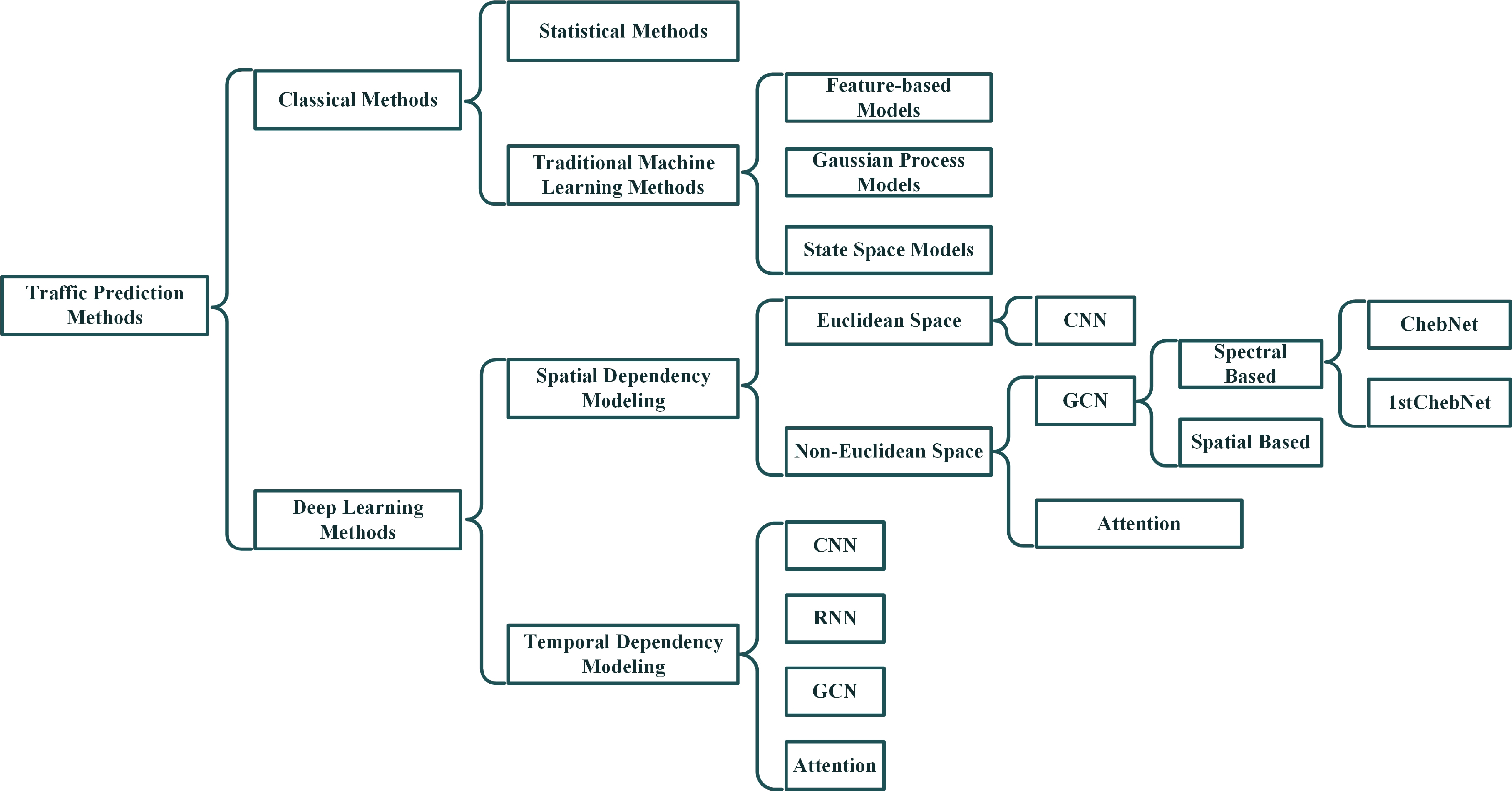}
\caption {Key techniques of traffic prediction methods.}
\label{framework}
\end{figure*}

It was not until the advent of deep learning-based methods that the full
potential of artificial intelligence in traffic prediction was developed
\cite{lv2015traffic}. This technology studies how to learn a hierarchical
model to map the original input directly to the expected output
\cite{goodfellow2016deep}. In general, deep learning models stack up basic
learnable blocks or layers to form a deep architecture, and the entire
network is trained end-to-end. Several architectures have been developed to
handle large-scale and complex spatio-temporal data. Generally, Convolutional
Neural Network (CNN) \cite{kalchbrenner2013recurrent} is employed to extract
spatial correlation of the grid-structured data described by images or
videos, and Graph Convolutional Network (GCN) \cite{bruna2014spectral}
extends convolution operation to more general graph-structured data, which is
more suitable to represent the traffic network structure. Furthermore,
Recurrent Neural Network (RNN)
\cite{rumelhart1988learning,elman1991distributed} and its variants LSTM
\cite{hochreiter1997long} or GRU \cite{cho2014learning} are commonly utilized
to model temporal dependency. Here, we summarize the key techniques commonly
used in existing traffic prediction methods, as shown in Fig.
\ref{framework}.

\textbf{Organization of this survey} The rest of this paper is organized as
follows. Section \ref{sec:traditional} covers the classical methods for
traffic prediction. Section \ref{sec:DLM} reviews the work based on deep
learning methods for traffic prediction, including the commonly used methods
of modeling spatial correlation and temporal correlation, as well as some
other new variants. Section \ref{sec:applications} lists some representative
results in each task. Section \ref{sec:Data} collects and organizes related
datasets and commonly used external data types for traffic prediction.
Section \ref{sec:experiments} provides some comparisons and evaluates the
performance of the relevant methods. Section \ref{sec:openproblem} discusses
several significant and important directions of future traffic prediction.
Finally, we conclude this paper in Section \ref{sec:con}.

\section{Classical Methods}
\label{sec:traditional} Statistical and traditional machine learning models
are two major representative data-driven methods for traffic prediction. In
time-series analysis, autoregressive integrated moving average
(ARIMA)\cite{williams2003modeling} and its variants are one of the most
consolidated approaches based on classical statistics and have been widely
applied for traffic prediction problems (\cite{
williams2003modeling,shekhar2007adaptive, li2012prediction,
moreira2013predicting, lippi2013short,wagner2017functional} ). However, these
methods are generally designed for small datasets, and are not suitable to
deal with complex and dynamic time series data. In addition, since usually
only temporal information is considered, the spatial dependency of traffic
data is ignored or barely considered.

Traditional machine learning methods, which can model more complex data, are
broadly divided into three categories: feature-based models, Gaussian process
models and state space models. Feature-based methods solve traffic prediction
problem (\cite{li2018general,guan2018a,tang2018forecasting} ) by training a
regression model based on human-engineered traffic features. These methods
are simple to implement and can provide predictions in some practical
situations. Gaussian process models the inner characteristics of traffic data
through different kernel functions, which need to contain spatial and
temporal correlations simultaneously. Although this kind of methods is proved
to be effective and feasible in traffic prediction
(\cite{diao2018hybrid,salinas2019high,lin2017road} ), compared to
feature-based models, they generally have higher computational load and
storage pressure, which is not appropriate when a mass of training samples
are available. State space models assume that the observations are generated
by Markovian hidden states. The advantage of this model is that it can
naturally model the uncertainty of the system and better capture the latent
structure of the spatio-temporal data. However, the overall non-linearity of
these models (\cite{duan2018unified,tan2016short,shin2018vehicle,
ishibashi2018inferring,gong2018network,polson2017bayesian,
hong2015learning,yu2016temporal,deng2016latent,deng2017situation,
kinoshita2016latent,Shvetsov2003Mathematical,Chiou2013Dynamic,gong2019potential,li2019tensor}
) is limited, and most of the time they are not optimal for modeling complex
and dynamic traffic data. Table \ref{traditional_methods} summarizes some
recent representative classical approaches.

\begin{table*}
\centering
\caption{Classical methods.}
\label{traditional_methods}
\begin{tabular}{c|c|c|c}
 \hline
 \multicolumn{2}{c|}{\rule{0pt}{12pt}Category}&Application task&Approach\\[5pt]
 \hline
 \hline
\multicolumn{2}{c|}{\multirow{2}{*}{Statistical methods}} &Flow &\cite{williams2003modeling,shekhar2007adaptive, lippi2013short,wagner2017functional}\\
\multicolumn{2}{c|}{}                                     &Demand &\cite{li2012prediction, moreira2013predicting}\\
 \hline
\multirow{11}{*}{Traditional machine learning methods} &\multirow{2}*{Feature-based models}      &Flow  &\cite{tang2018forecasting}\\
                                           &                                         &Demand&\cite{li2018general,guan2018a}\\
                                        \cline{2-4}
                                        &\multirow{4}*{Gaussian process models}      &Flow  &\cite{diao2018hybrid}\\
                                        &                                            &Speed &\cite{lin2017road}\\
                                        &                                            &Demand&\cite{salinas2019high}\\
                                        &                                            &Occupancy&\cite{salinas2019high}\\
                                        \cline{2-4}
                                        &\multirow{5}*{State space models}           &Flow  &\cite{tan2016short,polson2017bayesian,gong2018network,duan2018unified,gong2019potential,li2019tensor,hong2015learning,Shvetsov2003Mathematical,Chiou2013Dynamic}\\
                                        &                                            &Speed &\cite{deng2016latent,deng2017situation,shin2018vehicle}\\
                                        &                                            &Demand&\cite{ishibashi2018inferring}\\
                                        &                                            &Travel time&\cite{kinoshita2016latent}\\
                                        &                                            &Occupancy&\cite{kinoshita2016latent,yu2016temporal}\\

\hline
\end{tabular}
 \end{table*}

\section{Deep Learning Methods}
\label{sec:DLM}

Deep learning models exploit much more features and complex architectures
than the classical methods, and can achieve better performance. In Table
\ref{covered_literature}, we summarize the deep learning architectures in the
existing traffic prediction literature, and we will review these commonly
components in this section.

\subsection{Modeling Spatial Dependency}
\textbf{CNN.} A series of studies have applied CNN to capture spatial
correlations in traffic networks from two-dimensional spatio-temporal traffic
data\cite{Li2018a}. Since the traffic network is difficult to be described by
2D matrices, several researches try to convert the traffic network structure
at different times into images and divide these images into standard grids,
with each grid representing a region. In this way, CNNs can be used to learn
spatial features among different regions.
\begin{figure}[ht]
\centering
\includegraphics[width=0.5\linewidth]{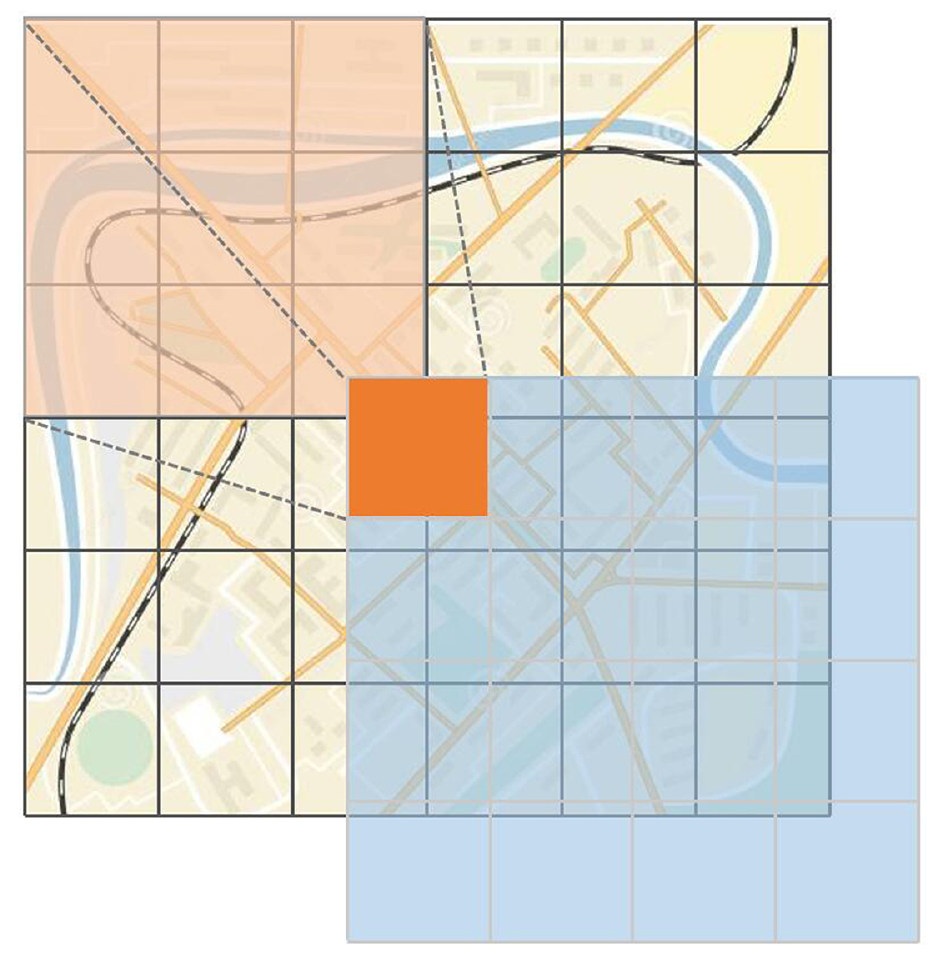}
\caption{2D Convolution. Each grid in the image is treated as a region, where neighbors are determined by the filter size. The 2D convolution operates between a certain region and its neighbors. The neighbors of the a region are ordered and have a fixed size.}
\label{cnn}
\end{figure}

As shown in Fig. \ref{cnn}, each region is directly connected to its nearby
regions. With a $3\times 3$ window, the neighborhood of each region is its
surrounding eight regions. The positions of these eight regions indicate an
ordering of a region's neighbors. A filter is then applied to this $3\times
3$ patch by taking the weighted average of the central region and its
neighbors across each channel. Due to the specific ordering of neighboring
regions, the trainable weights are able to be shared across different
locations.

In the division of traffic road network structure, there are many different
definitions of positions according to different granularity and semantic
meanings. \cite{zhang2016dnn} divided a city into I $\times$ J grid maps
based on the longitude and latitude where a grid represented a region. Then,
a CNN was applied to extract the spatial correlation between different
regions for traffic flow prediction.

\textbf{GCN.} Traditional CNN is limited to modeling Euclidean data, and GCN
is therefore used to model non-Euclidean spatial structure data, which is
more in line with the structure of traffic road network.  GCN generally
consists of two type of methods, spectral-based and spatial-based methods.
Spectral-based approaches define graph convolutions by introducing filters
from the perspective of graph signal processing where the graph convolution
operation is interpreted as removing noise from graph signals. Spatial-based
approaches formulate graph convolutions as aggregating feature information
from neighbors. In the following, we will introduce spectral-based GCNs and
spatial-based GCNs respectively.

\textit{(1) Spectral Methods.} Bruna et al. \cite{bruna2014spectral} first
developed spectral network, which performed convolution operation for graph
data from spectral domain by computing the eigen-decomposition of the graph
Laplacian matrix $\mathbf{L}$. Specifically, the graph convolution operation
$*_G$ of a signal $\mathbf{x}$ with a filter $\mathbf{g}\in{\mathbb{R}^{N}}$
can be defined as:
\begin{equation}\label{1}
\mathbf{x}*_G\mathbf{g}=\mathbf{U}\left(\mathbf{U}^T\mathbf{x}\odot \mathbf{U}^T\mathbf{g}\right),
\end{equation}
where $\mathbf{U}$ is the matrix of eigenvectors of normalized graph
Laplacian $\mathbf{L}$, which is defined as
$\mathbf{L}=\mathbf{I}_{N}-\mathbf{D}^{-\frac{1}{2}}\mathbf{A}\mathbf{D}^{-\frac{1}{2}}=\mathbf{U}\mathbf{\Lambda}\mathbf{U}^{T}$,
$\mathbf{D}$ is the diagonal matrix,
$\mathbf{D}_{ii}=\sum_{j}\left(\mathbf{A}_{ij}\right)$, $\mathbf{A}$ is the
adjacency matrix of the graph, $\mathbf{\Lambda}$ is the diagonal matrix of
eigenvalues, $\mathbf{\Lambda}=\lambda_{i}$. If we denote a filter as
$\mathbf{g}_{\theta}=diag\left(\mathbf{U}^{T}\mathbf{g}\right)$ parameterized
by $\theta\in\mathbb{R}^{N}$, the graph convolution can be simplified as:
\begin{equation}\label{2}
\mathbf{x}*_G\mathbf{g}=\mathbf{U}{\mathbf{g}_{\theta}}\mathbf{U}^{T}\mathbf{x},
\end{equation}
where a graph signal $\mathbf{x}$ is filtered by $\mathbf{g}$ with
multiplication between $\mathbf{g}$ and graph transform
$\mathbf{U}^{T}\mathbf{x}$. Though the computation of filter $\mathbf{g}$ in
graph convolution can be expensive due to $\mathcal{O}\left(n^2\right)$
multiplications with matrix $\mathbf{U}$, two approximation strategies have
been successively proposed to solve this issue.

\textit{ChebNet}. Defferrard et al. \cite{defferrard2016convolutional}
introduced a filter as Chebyshev polynomials of the diagonal matrix of
eigenvalues, i.e,
$\mathbf{g}_{\theta}=\sum_{i=1}^{K}\theta_{i}\mathbf{T}_{k}\left(\mathbf{\tilde{\Lambda}}\right)$,
where $\theta\in \mathbb{R}^{K}$ is now a vector of Chebyshev coefficients,
$\mathbf{\tilde{\Lambda}}=\frac{2}{\lambda_{max}}\mathbf{\Lambda}-\mathbf{I}_{N}$,
and $\lambda_{max}$ denotes the largest eigenvalue. The Chebyshev polynomials
are defined as
$\mathbf{T}_{k}(\mathbf{x})=2\mathbf{x}\mathbf{T}_{k-1}(\mathbf{x})-\mathbf{T}_{k-2}(\mathbf{x})$
with $\mathbf{T}_{0}{\mathbf{x}}=1$ and
$\mathbf{T}_{1}(\mathbf{x})=\mathbf{x}$. Then, the convolution operation of a
graph signal $\mathbf{x}$ with the defined filter $\mathbf{g}_{\theta}$ is:
\begin{equation}\label{3}
\begin{aligned}
\mathbf{x}\ast_{G}\mathbf{g}_{\theta}&=\mathbf{U}\left(\sum_{i=1}^{K}\theta_{i}\mathbf{T}_{k}\left(\mathbf{\tilde{\Lambda}}\right)\right)\mathbf{U}^{T}\mathbf{x}\\
                                    &=\sum_{i=1}^{K}\theta_{i}\mathbf{T}_{i}\left(\mathbf{\tilde{L}}\right)\mathbf{x},
\end{aligned}
\end{equation}
where  $\mathbf{\tilde{L}}=\frac{2}{\lambda_{max}}\mathbf{L}-\mathbf{I}_{N}$.

\textit{First order of ChebNet (1stChebNet)}. An first-order approximation of
ChebNet introduced by Kipf and Welling \cite{kipf2017semi} further simplified
the filtering by assuming $K=1$ and $\lambda_{max}=2$, we can obtain the
following simplified expression:
\begin{equation}\label{4}
\mathbf{x}\ast_{G}\mathbf{g}_{\theta}=\theta_{0}\mathbf{x}-\theta_{1}\mathbf{D}^{-\frac{1}{2}}\mathbf{A}\mathbf{D}^{-\frac{1}{2}}\mathbf{x},
\end{equation}
where $\theta_{0}$ and $\theta_{1}$ are learnable parameters. After further
assuming these two free parameters with $\theta=\theta_{0}=-\theta_{1}$. This
can be obtained equivalently in the following matrix form:
\begin{equation}\label{5}
\mathbf{x}\ast_{G}\mathbf{g}_{\theta}=\theta\left(\mathbf{I}_{N}+\mathbf{D}^{-\frac{1}{2}}\mathbf{A}\mathbf{D}^{-\frac{1}{2}}\right)\mathbf{x}.
\end{equation}

To avoid numerical instabilities and exploding/vanishing gradients due to stack operations, another normalization technique is introduced: $\mathbf{I}_{N}+\mathbf{D}^{-\frac{1}{2}}\mathbf{A}\mathbf{D}^{-\frac{1}{2}}\rightarrow\tilde{\mathbf{D}}^{-\frac{1}{2}}\tilde{\mathbf{A}}\tilde{\mathbf{D}}^{-\frac{1}{2}}$, with $\tilde{\mathbf{A}}=\mathbf{A}+\mathbf{I}_{N}$ and $\tilde{\mathbf{D}}_{ii}=\sum_{j}\tilde{\mathbf{A}}_{ij}$.
Finally, a graph convolution operation can be changed to:
\begin{equation}\label{6}
\mathbf{Z}=\tilde{\mathbf{D}}^{-\frac{1}{2}}\tilde{\mathbf{A}}\tilde{\mathbf{D}}^{-\frac{1}{2}}\mathbf{X}
\mathbf{\Theta},
\end{equation}
where $\mathbf{X}\in\mathbb{R}^{N\times C}$ is a signal, $\mathbf{\Theta}\in\mathbb{R}^{C\times F}$ is a matrix of filter parameters, $C$ is the input channels, $F$ is the number of filters, and $Z$ is the transformed signal matrix.

To fully utilize spatial information, \cite{yu2017spatio} modeled the traffic
network as a general graph rather than treating it as grids, where the
monitoring stations in a traffic network represent the nodes in the graph,
the connections between stations represent the edges, and the adjacency
matrix is computed based on the distances among stations, which is a natural
and reasonable way to formulate the road network. Afterwards, two graph
convolution approximation strategies based on spectral methods were used to
extract patterns and features in the spatial domain, and the computational
complexity was also reduced. \cite{geng2019spatiotemporal} first used graphs
to encode different kinds of correlations among regions, including
neighborhood, functional similarity, and transportation connectivity. Then,
three groups of GCN based on ChebNet were used to model spatial correlations
respectively, and traffic demand prediction was made after further
integrating temporal information.

\begin{figure}[ht]
\centering
\includegraphics[width=0.5\linewidth]{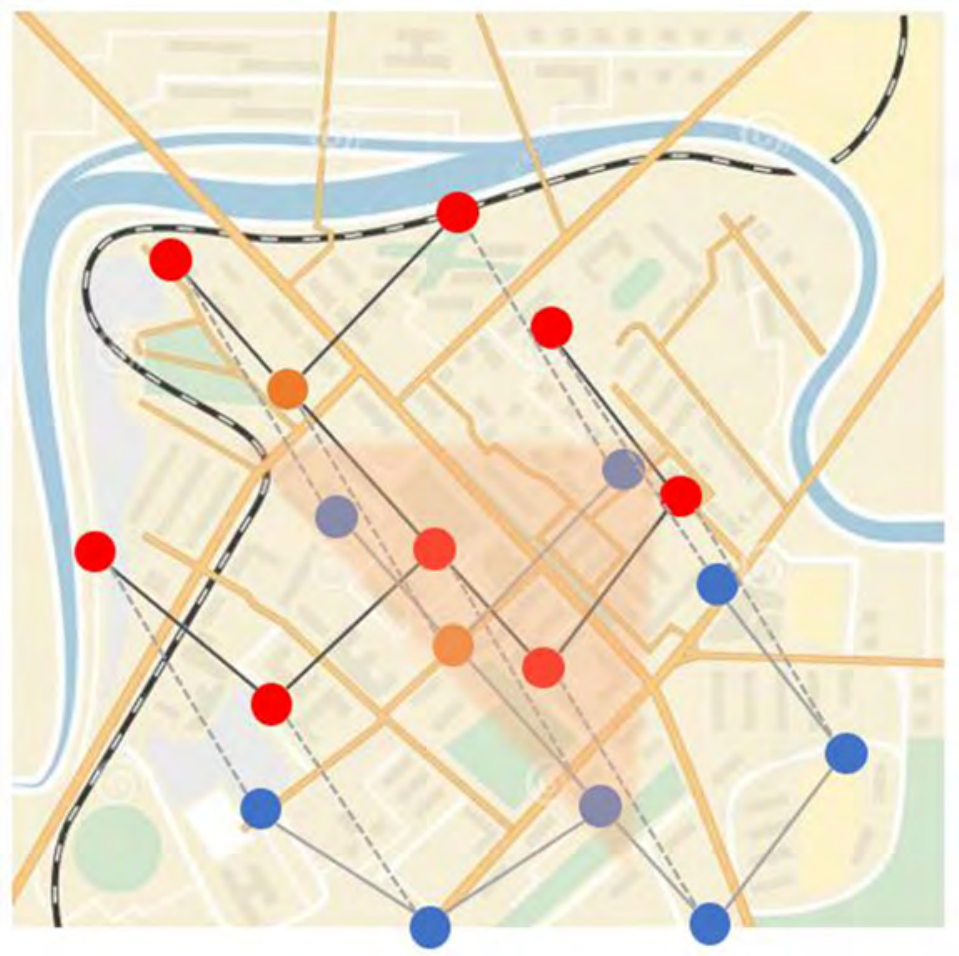}
\caption{Spatial-based graph convolution network. Each node in the graph can represent a region in the traffic network.
To get a hidden representation of a certain node (e.g. the orange node), GCN aggregates feature information from its neighbors (shaded area).
Unlike grid data in 2D images, the neighbors of a region are unordered and varies in size.}
\label{gcn}
\end{figure}

\textit{(2) Spatial Methods.} Spatial methods define convolutions directly on
the graph through the aggregation process that operates on the central node
and its neighbors to obtain a new representation of the central node, as
depicted by Fig.\ref{gcn}. In \cite{li2018diffusion}, traffic network was
firstly modeled as a directed graph, the dynamics of the traffic flow was
captured based on the diffusion process. Then a diffusion convolution
operation is applied to model the spatial correlation, which is a more
intuitive interpretation and proves to be effective in spatial-temporal
modeling. Specifically, diffusion convolution models the bidirectional
diffusion process, enabling the model to capture the influence of upstream
and downstream traffic. This process can be defined as:
\begin{equation}\label{4}
\mathbf{X}_{:,p}\star_{G}f_{\theta}=\sum_{k=0}^{K-1}\left(\theta_{k1}\left(\mathbf{D}_{O}^{-1}\mathbf{A}\right)^{k}
+\theta_{k2}(\mathbf{D}_{I}^{-1}\mathbf{A}^{T})^k\right)\mathbf{X}_{:,p},
\end{equation}
where $\mathbf{X}\in \mathbb{R}^{N\times P}$ is the input, $P$ represents the
number of input features of each node. $\star_{G}$ denotes the diffusion
convolution, $k$ is the diffusion step, $f_{\theta}$ is a filter and
$\theta\in\mathbb{R}^{K\times 2}$ are learnable parameters. $\mathbf{D}_{O}$
and $\mathbf{D}_{I}$ are out-degree and in-degree matrices respectively. To
allow multiple input and output channels, DCRNN \cite{li2018diffusion}
proposes a diffusion convolution layer, defined as:
\begin{equation}\label{4}
\mathbf{Z}_{:,p}=\sigma\left(\sum_{p=1}^{P}\mathbf{X}_{:,p}\star_{G}f\mathbf{\Theta}_{q,p,:,:}\right),
\end{equation}
where $\mathbf{Z}\in \mathbb{R}^{N\times Q}$ is the output,
$\mathbf{\Theta}\in \mathbb{R}^{Q\times P\times K\times 2}$ parameterizes the
convolutional filter , $Q$ is the number of output features, $\sigma$ is the
activation function. Based on the diffusion convolution process,
\cite{chen2019gated} designed a new neural network layer that can map the
transformation of different dimensional features and extract patterns and
features in spatial domain. \cite{wu2019graph} modified the diffusion process
in \cite{li2018diffusion} by utilizing a self-adaptive adjacency matrix,
which allowed the model to mine hidden spatial dependency by itself.
\cite{Song2020Spatial} introduced the notion of aggregation to define graph
convolution. This operation can assemble the features of each node with its
neighbors. The aggregate function is a linear combination whose weights are
equal to the weights of the edges between the node and its neighbors. This
graph convolutional operation can be expressed as follow:
\begin{equation}\label{4}
h^{(l)}=\sigma(\mathbf{A}h^{(l-1)}W+b),
\end{equation}
where $h^{(l-1)}$ is the input of the $l$-th graph convolutional layer, $W$
and $b$ are parameters, and $\sigma$ is the activation function.

\textbf{Attention.} Attention mechanism is first proposed for natural
language processing\cite{bahdanau2014neural}, and has been widely used in
various fields. The traffic condition of a road is affected by other roads
with different impacts. Such impact is highly dynamic, changing over time. To
model these properties, the spatial attention mechanism is often used to
adaptively capture the correlations between regions in the road network
(\cite{pan2019urban,li2019learning,zhang2018gaan,li2019forecaster,zheng2019gman,park2019stgrat,geng2020cgt,9062547,yin2021multi}
). The key idea is to dynamically assign different weights to different
regions at different time steps. For the sake of simplicity, we ignore time
coordinates for the moment. Attention mechanism operates on a set of input
sequence $x=(x_{1},\ldots,x_{n})$ with $n$ elements where
$x_{i}\in\mathbb{R}^{d_{x}}$, and computes a new sequence
$z=(z_{1},\ldots,z_{n})$ with the same length where
$z_{i}\in\mathbb{R}^{d_{z}}$. Each output element $z_{i}$ is computed as a
weighted sum of a linear transformed input elements:
\begin{equation}\label{7}
z_{i}=\sum_{j=1}^{n}\alpha_{ij}x_{j}.
\end{equation}

The weight coefficient $\alpha_{ij}$ indicates the importance of $x_{i}$ to
$x_{j}$, and it is computed by a softmax function:
\begin{equation}\label{8}
\alpha_{ij}=\frac{\exp e_{ij}}{\sum_{k=1}^{n}\exp e_{ik}},
\end{equation}
where $e_{ij}$ is computed using a compatibility function that compares two
input elements:
\begin{equation}\label{9}
e_{ij}=v^{\top}tanh \left(x_{i}W^{Q}+x_{j}W^{k}+b\right),
\end{equation}
and generally Perceptron is chosen for the compatibility function. Here, the
learnable parameters are $v$, $W^{Q}$, $W^{k}$ and $b$. This mechanism has
proven effective, but when the number of elements $n$ in a sequence is large,
we need to calculate $n^2$ weight coefficients, and therefore the time and
memory consumption are heavy.

In traffic speed prediction, \cite{zhang2018gaan} used attention mechanism to
dynamically capture the spatial correlation between the target region and the
first-order neighboring regions of the road network. \cite{guo2019attention}
combined the GCN based on ChebNet with attention mechanism to make full use
of the topological properties of the traffic network and dynamically adjust
the correlations between different regions.

\subsection{Modeling Temporal Dependency}
\textbf{CNN.}  \cite{gehring2017convolutional} first introduced the fully
convolutional model for sequence to sequence learning. A representative work
in traffic research, \cite{yu2017spatio} applied purely  convolutional
structures to simultaneously extract spatio-temporal features from
graph-structured time series data. In addition, dilated causal convolution is
a special kind of standard one-dimensional convolution. It adjusts the size
of the receptive field by changing the value of the dilation rate, which is
conducive to capture the long-term periodic dependence. \cite{fang2019gstnet}
and \cite{yao2019learning} therefore adopted the dilated causal convolution
as the temporal convolution layer of their models to capture a node's
temporal trends. Compared to recurrent models, convolutions create
representations for fixed size contexts, however, the effective context size
of the network can easily be made larger by stacking several layers on top of
each other. This allows to precisely control the maximum length of
dependencies to be modeled. The convolutional network does not rely on the
calculation of the previous time step, so it allows parallelization of every
element in the sequence, which can make better use of GPU hardware, and
easier to optimize. This is superior to RNNs, which maintain the entire
hidden state of the past, preventing parallel calculations in a sequence.

\textbf{RNN.} RNN and its variant LSTM or GRU, are neural networks for
processing sequential data. To model the non-linear temporal dependency of
traffic data, RNN-based approaches have been applied to traffic
prediction\cite{Li2018a}. These models rely on the order of data to process
data in turn, and therefore one disadvantage of these models is that when
modeling long sequences, their ability to remember what they learned before
many time steps may decline.

In RNN-based sequence learning, a special network structure known as
encoder-decoder has been applied for traffic prediction (
\cite{zhu2017deep,li2019forecaster,ye2019co,li2018diffusion,liao2018deep,jiang2019deepurbanevent,zheng2019gman,zhang2018multistep,
wang2019origin,bai2019stg2seq,park2019stgrat,deshpande2019streaming,
pan2019urban,chai2018bike,geng2020cgt,9062547,yin2021multi} ). The key idea
is to encode the source sequence as a fixed-length vector and use the decoder
to generate the prediction.
\begin{equation}\label{10}
\mathbf{s}=f\left(\mathcal{F}_{t};\theta_1\right),
\end{equation}
\begin{equation}\label{11}
\mathbf{\hat{X}}_{t+1:t+L}=g\left(\mathbf{s};\theta_2\right),
\end{equation}
where $f$ is the encoder and $g$ is the decoder. $\mathcal{F}_{t}$ denotes
the input information available at timestamp $t$, $\mathbf{s}$ is a
transformed semantic vector representation, $\mathbf{\hat{X}}_{t+1:t+L}$ is
the value of $L$-step-ahead prediction, $\theta_1$ and $\theta_2$ are
learning parameters.

One potential problem with encoder-decoder structure is that regardless of
the length of the input and output sequences, the length of semantic vector
$\mathbf{s}$ between encoding and decoding is always fixed, and therefore
when the input information is too long, some information will be lost.

\textbf{Attention.} To resolve the above issue, an important extension is
to use an attention mechanism on time axis, which can adaptively select the
relevant hidden states of the encoder to produce output sequence. This is
similar to attention in the spatial methods. Such a temporal attention
mechanism can not only model the non-linear correlation between the current
traffic condition and the previous observations at a certain position in the
road network, but also model the long-term sequence data to solve the
deficiencies of RNN.

\cite{zheng2019gman} designed a temporal attention mechanism to adaptively
model the non-linear correlations between different time steps.
\cite{guo2019attention} incorporated a standard convolution and attention
mechanism to update the information of a node by fusing the information at
the neighboring time steps, and semantically express the dependency intensity
between different time steps. Considering that traffic data is highly
periodic, but not strictly periodic, \cite{yao2019revisiting} designed a
periodically shifted attention mechanism to deal with long-term periodic
dependency and periodic temporal shifting.

\textbf{GCN.} Song et al. first constructed a localized spatio-temporal graph
that includes both temporal and spatial attributes, and then used the
proposed spatial-based GCN method to model the spatio-temporal correlations
simultaneously \cite{Song2020Spatial}.

\subsection{Joint Spatio-Temporal Relationships Modeling}
As shown in Table \ref{covered_literature}, most methods use a hybrid deep
learning framework, which combines different types of techniques to capture
the spatial dependencies and temporal correlations of traffic data
separately. They assume that the relations of geographic information and
temporal information are independent and do not consider their joint
relations. Therefore, the spatial and temporal correlations are not fully
exploited to obtain better accuracy. To solve this limitation, researchers
have attempted to integrate spatial and temporal information into an
adjacency graph matrix or tensor. For example, \cite{Song2020Spatial} got a
localized spatio-temporal graph by connecting all nodes with themselves at
the previous moment and the next moment. According to the topological
structure of the localized spatial-temporal graph, the correlations between
each node and its spatio-temporal neighbors can be captured directly. In
\cite{fang2020constgat}, Fang et al. constructed three matrices for the
historical traffic conditions of different links, the features of the
neighbor links, and the features of the historical time slots, in which each
row of the matrix corresponds to the information of a link. Finally, these
three matrices were concatenated into a matrix and reshaped into a 3D
spatio-temporal tensor. Attention mechanism was then used to obtain the
relations between the traffic conditions.

\subsection{Deep Learning plus Classical Models}
Recently, more and more researches are combining deep learning with classical
methods, and some advanced methods have been used in traffic prediction
(\cite{li2019learninginter,pan2019matrix,sen2019think,ziat2017spatio} ). This
kind of method not only makes up for the weak ability of non-linear
representation of classical models but also makes up for the poor
interpretability of deep learning methods. \cite{li2019learninginter}
proposed a method based on the generation model of state space and the
inference model based on filtering, using deep neural networks to realize the
non-linearity of the emission and the transition models, and using the
recurrent neural network to realize the dependence over time. Such a
non-linear network based parameterization provides the flexibility to deal
with arbitrary data distribution. \cite{pan2019matrix} proposed a deep
learning framework that introduced matrix factorization method into deep
learning model, which can model the latent region functions along with the
correlations among regions, and further improve the model capability of the
citywide flow prediction. \cite{sen2019think} developed a hybrid model that
associated a global matrix decomposition model regularized by a temporal deep
network with a local deep temporal model that captured patterns specific to
each dimension. Global and local models are combined through a data-driven
attention mechanism for each dimension. Therefore, global patterns of the
data can be utilized and combined with local calibration for better
prediction. \cite{ziat2017spatio} combined a latent model and multi-layer
perceptrons (MLP) to design a network for addressing multivariate
spatio-temporal time series prediction problems. The model captures the
dynamics and correlations of multiple series at the spatial and temporal
levels. Table \ref{covered_plus_literature} summarizes relevant literatures
in terms of deep learning plus classical methods.

\subsection{Limitations of the Deep Learning-based Method} The
strengths of the deep neural network model make it very attractive and indeed
greatly promote the progress in the field of traffic prediction. However, it
also possess several disadvantages compared with classical methods.
\begin{itemize}
\item High data demand. Deep learning is highly data-dependent, and
    typically the larger the amount of data, the better it performs. In
    many cases, such data is not readily available, for example, some
    cities may release taxi data for multiple years, while others release
    data for just a few days.
\item High computational complexity. Deep learning requires high computing
    power, and ordinary CPUs can no longer meet the requirements of deep
    learning. The mainstream computing uses GPU and TPU. At the same time,
    with the increase of model complexity and the number of parameters, the
    demand for memory is also gradually increasing. In general, deep neural
    networks are more computationally expensive than classical algorithms.
\item Lack of interpretability. Deep learning models are mostly considered
    as ``black-boxs" that lack interpretability. In general, the prediction
    accuracy of deep learning models is higher than that of classical
    methods. However, there is no explanation as to why these results are
    obtained or how parameters can be determined to make the results
    better.

\end{itemize}
\section{Representative Results}
\label{sec:applications}

\begin{table*}
\centering
\caption{Categorization for the covered deep learning literature.}
\label{covered_literature}
\begin{tabular}{c|c|c|c|m{6cm}<{\centering}}
 \hline
\rule{0pt}{12pt} Application task&Spatial modeling type&\multicolumn{2}{c|}{Temporal modeling type}&Approach\\[5pt]
 \hline
 \hline
 \multirow{10}{*}{Flow}&{\multirow{2}{*}{CNN}}&\multicolumn{2}{c|}{--}&\cite{lin2019deepstn+,zhang2016dnn,zhang2017deep,zhang2019flow,guo2019deep,li2020autost}\\\cline{3-5}
                       &&\multicolumn{2}{c|}{RNN}&\cite{yao2019learning,wang2018cross,zhao2018layerwise,zonoozi2018periodic,jiang2019deepurbanevent,zheng2019deep}\\\cline{2-5}
                       &1stChebNet&\multicolumn{2}{c|}{RNN}&\cite{chai2018bike,bai2020adaptive}\\\cline{2-5}
                       &\multirow{2}{*}{ChebNet}&\multicolumn{2}{c|}{CNN (Causal CNN)}&\cite{fang2019gstnet}\\\cline{3-5}
                       &&\multicolumn{2}{c|}{RNN}&\cite{guo2020optimized}\\\cline{2-5}
                       &GCN+Attention&\multicolumn{2}{c|}{CNN (1-D Conv) +Attention}&\cite{guo2019attention}\\\cline{2-5}
                       &{\multirow{4}{*}{Attention only}}&\multicolumn{2}{c|}{--}&\cite{liu2020learning}\\\cline{3-5}
                       &              &\multicolumn{2}{c|}{RNN}&\cite{pan2019urban}\\\cline{3-5}
                       &              &\multicolumn{2}{c|}{RNN+Attention}&\cite{9062547,yin2021multi}\\\cline{3-5}
                       &              &\multicolumn{2}{c|}{Attention only}&\cite{zheng2019gman}\\

\hline
 \multirow{14}{*}{Speed}&--&\multicolumn{2}{c|}{RNN}&\cite{wang2018efficient,tang2019joint}\\\cline{2-5}
                      &CNN&\multicolumn{2}{c|}{RNN}&\cite{zang2018long,lv2018lc}\\\cline{2-5}
                      &\multirow{2}{*}{1stChebNet}&\multicolumn{2}{c|}{CNN (1-D Conv)}&\cite{yu2017spatio}\\\cline{3-5}
                      &&\multicolumn{2}{c|}{RNN}&\cite{zhao2019t}\\\cline{2-5}
                      &\multirow{3}{*}{ChebNet}&\multirow{2}{*}{CNN}&(1-D Conv)&\cite{yu2017spatio}\\\cline{4-5}
                      &                        &                    &(2-D Conv)&\cite{diao2019dynamic}\\\cline{3-5}
                      &&\multicolumn{2}{c|}{RNN}&\cite{liao2018deep,zhang2019hybrid,guo2020optimized,cui2019traffic}\\\cline{3-5}
                      &&\multicolumn{2}{c|}{RNN+Attention}&\cite{zhang2018multistep}\\\cline{2-5}
                      &\multirow{2}{*}{GCN(spatial-based)}&\multicolumn{2}{c|}{CNN (Causal CNN)}&\cite{wu2019graph}\\\cline{3-5}
                      &&\multicolumn{2}{c|}{RNN}&\cite{li2018diffusion,chen2019gated}\\\cline{2-5}
                      &\multirow{2}{*}{GCN+Attention}&\multicolumn{2}{c|}{CNN (1-D Conv)}&\cite{huang2020long}\\\cline{3-5}
                      &&\multicolumn{2}{c|}{RNN}&\cite{chen2020multi}\\\cline{2-5}
                      &\multirow{2}{*}{Attention only}&\multicolumn{2}{c|}{RNN}&\cite{zhang2018gaan,pan2019urban}\\\cline{3-5}
                      &&\multicolumn{2}{c|}{Attention only}&\cite{zheng2019gman,park2019stgrat}\\

 \hline
 \multirow{9}{*}{Demand}
                       &--&\multicolumn{2}{c|}{RNN}&\cite{xu2017real}\\\cline{2-5}
                       &\multirow{3}{*}{CNN}&\multicolumn{2}{c|}{--}&\cite{lee2018forecasting}\\\cline{3-5}
                       &&\multicolumn{2}{c|}{RNN}&\cite{liu2019contextualized,ye2019co,yao2018deep,ke2017short}\\\cline{3-5}
                       &&\multicolumn{2}{c|}{RNN+Attention}&\cite{yao2019revisiting}\\\cline{2-5}
                       &\multirow{2}{*}{1stChebNet}&\multicolumn{2}{c|}{Attention only}&\cite{bai2019stg2seq}\\\cline{3-5}
                       &&\multicolumn{2}{c|}{RNN}&\cite{wang2019origin,davis2019grids}\\\cline{2-5}
                       &ChebNet&\multicolumn{2}{c|}{RNN}&\cite{geng2019spatiotemporal}\\\cline{2-5}
                       &\multirow{2}{*}{Attention only}&\multicolumn{2}{c|}{RNN}&\cite{li2019learning}\\\cline{3-5}
                       &&\multicolumn{2}{c|}{Attention only}&\cite{geng2020cgt,li2019forecaster}\\

 \hline
 \multirow{3}{*}{Travel time}&--&\multicolumn{2}{c|}{RNN}&\cite{pang2018learning,he2018travel}\\\cline{2-5}
                             &CNN&\multicolumn{2}{c|}{RNN}&\cite{wang2018will}\\\cline{2-5}
                             &ChebNet&\multicolumn{2}{c|}{CNN (1-D Conv)}&\cite{dai2020hybrid}\\
 \hline
 \multirow{2}{*}{Occupancy}&--&\multicolumn{2}{c|}{RNN}&\cite{deshpande2019streaming}\\\cline{2-5}
                           &CNN&\multicolumn{2}{c|}{RNN}&\cite{lai2018modeling}\\
 \hline
 \end{tabular}
 \end{table*}

\begin{table*}
\centering
\caption{Categorization for the covered deep learning plus classical literature.}
\label{covered_plus_literature}
\begin{tabular}{c|c|c}
 \hline
\rule{0pt}{12pt} Application task&Approach&Spatio-temporal modeling\\[5pt]
 \hline
 \hline
 \multirow{2}{*}{Flow}&\cite{ziat2017spatio}&State space model+MLP\\\cline{2-3}
                      &\cite{pan2019matrix}&State space model+CNN+RNN\\
 \hline
 Demand&\cite{pan2019matrix}&State space model+CNN+RNN\\
 \hline
 \multirow{2}{*}{Occupancy}&\cite{li2019learninginter}&State space model+RNN\\\cline{2-3}

 &\cite{sen2019think}&State space model+CNN\\
 \hline
 \end{tabular}
 \end{table*}

In this section, we summarize some representative results of different
application tasks. Based on the literature studied on different tasks, we
list the current best performance methods under commonly used public
datasets, as shown in Table \ref{prediction_performance}. We can have the
following observations: First, the results on different datasets vary greatly
under the same prediction task. For example, in the demand prediction task,
the NYC Taxi and TaxiBJ datasets obtained the accuracy of 8.385 and 17.24,
respectively, under the same time interval and prediction time. Under the
same condition of the prediction task and the dataset, the performance
decreases with the increase of prediction time, as shown in the speed
prediction results on Q-Traffic. For the dataset of the same data source, due
to the different time and region selected, it also has a greater impact on
the accuracy, e.g., related datasets based on PeMS under the speed prediction
task. Second, in different prediction tasks, the accuracy of speed prediction
task can reach above 90\% in general, which is significantly higher than
other tasks whose accuracy rate is close to or more than 80\%. Therefore,
there is still much room for improvement in these tasks.
\begin{table*}
\centering
\caption{Prediction performance statistics for different tasks.}
\label{prediction_performance}
\begin{tabular}{c|c|c|c|c|c}
 \hline
 \rule{0pt}{12pt}Application task&Dataset&Time interval&Prediction window&MAPE&RMSE\\[5pt]
 \hline
 \hline
 \multirow{6}*{Flow}& TaxiBJ   &   30min & 30min & 25.97\%\cite{li2020autost}    & 15.88\cite{li2020autost}\\\cline{2-6}
 & PeMSD3 & 5min & 60min & 16.78\%\cite{Song2020Spatial} & 29.21\cite{Song2020Spatial}\\\cline{2-6}
                    & PeMSD4   &   5min & 60min & 11.09\%\cite{9062547}    & 31.00\cite{9062547}\\\cline{2-6}
                    & PEMS07   &   5min & 60min & 10.21\%\cite{Song2020Spatial} & 38.58\cite{li2018diffusion}\\\cline{2-6}
                    & PeMSD8   &   5min & 60min & 8.31\%\cite{9062547}    & 24.74\cite{9062547}\\\cline{2-6}
                    & NYC Bike &   60min & 60min & --    & 6.33\cite{zhang2017deep}\\\cline{2-6}
                    & T-Drive  &   60min & 60/120/180min & -- & 29.9/34.7/37.1\cite{pan2019urban}\\
\hline
 \multirow{7}*{Speed} & METR-LA  &   5min & 5/15/30/60min &4.90\%\cite{chen2019gated}/6.80\%/8.30\%/10.00\%\cite{chen2020multi} & 3.57\cite{chen2019gated}/5.12/6.17/7.30\cite{chen2020multi}\\\cline{2-6}
                      & PeMS-BAY &   5min & 15/30/60min   &2.73\%\cite{wu2019graph}/3.63\%\cite{zheng2019gman}/4.31\%\cite{zheng2019gman}  & 2.74\cite{wu2019graph}/3.70\cite{wu2019graph}/4.32\cite{zheng2019gman}\\\cline{2-6}
                      & PeMSD4   &   5min & 15/30/45/60min &2.68\%\cite{li2018diffusion}/3.71\%\cite{li2018diffusion}/4.42\%/4.85\%\cite{huang2020long}&2.93/3.92/4.47/4.83\cite{huang2020long}\\\cline{2-6}
                      & PeMSD7   &   5min & 15/30/45/60min &5.14\%/7.18\%/8.51\%/9.60\%\cite{huang2020long}&3.98/5.47/6.39/7.09\cite{huang2020long}\\\cline{2-6}
                      & PeMSD7(M) & 5min & 15/30/45min & 5.24\%/7.33\%/8.69\%\cite{yu2017spatio} & 4.04/5.70/6.77\cite{yu2017spatio}\\\cline{2-6}
                      & PeMSD8   &   5min & 15/30/45/60min &2.24\%/3.02\%/3.51\%/3.89\%\cite{huang2020long}&2.45/3.28/3.75/4.11\cite{huang2020long}\\\cline{2-6}
                      & SZ-taxi & 15min & 15/30/45/60min & -- &3.92/3.96/3.98/4.00\cite{zhao2019t}\\\cline{2-6}
                      & Los-loop & 5min & 15/30/45/60min & --& 5.12/6.05/6.70/7.26\cite{zhao2019t}\\\cline{2-6}
                      & LOOP & 5min & 5min & 6.01\%\cite{cui2019traffic} & 4.63\cite{cui2019traffic}\\\cline{2-6}
                      & Q-Traffic & 15min &  \makecell[c]{15/30/45/60/\\ 75/90/105/120min} & \makecell[c]{4.52\%/7.93\%/8.89\%/9.24\%/\\9.43\%/9.56\%/9.69\%/9.78\% }\cite{liao2018deep}& --\\\cline{2-6}

 \hline
\multirow{3}*{Demand}& NYC Taxi  &   30min & 30min & -- & 8.38\cite{ye2019co}\\\cline{2-6}
                      & NYC Bike  &   60min & 60min & 21.00\%\cite{bai2019stg2seq} &4.51\cite{bai2019stg2seq}\\\cline{2-6}
                      & TaxiBJ    &   30min & 30min & 13.80\%\cite{bai2019stg2seq} &17.24\cite{bai2019stg2seq}\\\cline{2-6}

 \hline
Travel time&Chengdu&--&--&11.89\%\cite{wang2018will}&--\\
\hline
Occupancy&PeMSD-SF&60min&\makecell[c]{7 rolling time windows\\(24 time-points at a time)}&16.80\%\cite{sen2019think}&--\\
\hline
 \end{tabular}
 \end{table*}

Some companies are currently conducting intelligent transportation research,
such as amap, DiDi, and Baidu maps. According to amap technology annual in
2019\cite{alicdn}, amap has carried out the exploration and practice of deep
learning in the prediction of the historical speed of amap driving
navigation, which is different from the common historical average method and
takes into account the timeliness and annual periodicity characteristics
presented in the historical data. By introducing the Temporal Convolutional
Network (TCN) \cite{lea2017temporal} model for industrial practice, and
combining feature engineering (extracting dynamic and static features,
introducing annual periodicity, etc.), the shortcomings of existing models
are successfully solved. The arrival time of a given week is measured based
on the order data, and it has a badcase rate of 10.1\%, which is 0.9\% lower
than the baseline. For the travel time prediction in the next hour,
\cite{dai2020hybrid} designed a multi-model architecture to infer the future
travel time by adding contextual information using the upcoming traffic flow
data. Using anonymous user data from amap, MAPE can be reduced to around 16\%
in Beijing.

The Estimated Time of Arrival (ETA), supply and demand and speed prediction
are the key technologies in DiDi's platform. DiDi has applied artificial
intelligence technology in ETA, reduced MAPE index to 11\% by utilizing
neural network and DiDi's massive order data, and realized the ability to
provide users with accurate expectation of arrival time and multi-strategy
path planning under real-time large-scale requests. In the prediction and
scheduling, DiDi has used deep learning model to predict the difference
between supply and demand after some time in the future, and provided driver
scheduling service. The prediction accuracy of the gap between supply and
demand in the next 30 minutes has reached 85\%. In the urban road speed
prediction task, DiDi proposed a prediction model based on driving trajectory
calibration\cite{zhang2019boosted}. Through comparison experiments based on
Chengdu and Xi'an data in the DiDi gaia dataset, it was concluded that the
overall MSE indicator for speed prediction was reduced to 3.8 and 3.4.

Baidu has solved the traffic prediction task of online route queries by
integrating auxiliary information into deep learning technology, and released
a large-scale traffic prediction dataset from Baidu Map with offline and
online auxiliary information\cite{liao2018deep}. The overall MAPE and 2-hour
MAPE of speed prediction on this dataset decreased to 8.63\% and 9.78\%,
respectively. In \cite{fang2020constgat}, the researchers proposed an
end-to-end neural framework as an industrial solution for the travel time
prediction function in mobile map applications, aiming at exploration of
spatio-temporal relation and contextual information in traffic prediction.
The MAPE in Taiyuan, Hefei and Huizhou, sampled on the Baidu maps, can be
reduced to 21.79\%, 25.99\% and 27.10\% respectively, which proves the
superiority of the model. The model is already in production on Baidu maps
and successfully handles tens of billions of requests a day.

\section{Public datasets}
\label{sec:Data} High-quality datasets are essential for accurate traffic
forecasting. In this section, we comprehensively summarize the public data
information used for the prediction task, which mainly consists of two parts:
one is the public spatio-temporal sequence data commonly used in the
prediction, and the other is the external data to improve the prediction
accuracy. However, the latter data is not used by all models due to the
design of different model frameworks or the availability of the data.

\textbf{Public datasets} Here, we list public, commonly used and large-scale
real-world datasets in traffic prediction.

\begin{itemize}
\item PeMS: It is an abbreviation from the California Transportation Agency
    Performance Measurement System (PeMS), which is displayed on the map
    and collected in real-time by more than 39000 independent detectors.
    These sensors span the freeway system across all major metropolitan
    areas of the State of California. The source is available at:
    http://pems.dot.ca.gov/. Based on this system, several sub-dataset
    versions (PeMSD3/4/7(M)/7/8/-SF/-BAY) have appeared and are widely
    used. The main difference is the range of time and space, as well as
    the number of sensors included in the data collection.

\textit{PeMSD3}: This dataset is a piece of data processed by Song et al.
It includes 358 sensors and flow information from 9/1/2018 to 11/30/2018. A
processed version is available at: https://github.com/Davidham3/STSGCN.

\textit{PeMSD4}: It describes the San Francisco Bay Area, and contains 3848
sensors on 29 roads dated from 1/1/2018 until 2/28/2018, 59 days in total.
A processed version is available at:
https://github.com/Davidham3/ASTGCN/tree/master/data/ PEMS04.

\textit{PeMSD7(M)}: It describes the District 7 of California containing
228 stations, and The time range of it is in the weekdays of May and June
of 2012. A processed version is available at:
https://github.com/Davidham3/STGCN/tree/master/ datasets.

\textit{PeMSD7}: This version was publicly released by Song et al. It
contains traffic flow information from 883 sensor stations, covering the
period from 7/1/2016 to 8/31/2016. A processed version is available at:
https://github.com/Davidham3/STSGCN.

\textit{PeMSD8}: It depicts the San Bernardino Area, and contains 1979
sensors on 8 roads dated from 7/1/2016 until 8/31/2016, 62 days in total. A
processed version is available at:
https://github.com/Davidham3/ASTGCN/tree/master/ data/PEMS08.

\textit{PeMSD-SF}: This dataset describes the occupancy rate, between 0 and
1, of different car lanes of San Francisco bay area freeways. The time span
of these measurements is from 1/1/2008 to 3/30/2009 and the data is sampled
every 10 minutes. The source is available at:
http://archive.ics.uci.edu/ml/datasets/PEMS-SF.

\textit{PeMSD-BAY}: It contains 6 months of statistics on traffic speed,
ranging from 1/1/2017 to 6/30/2017, including 325 sensors in the Bay area.
The source is available at: https://github.com/liyaguang/DCRNN.

\item METR-LA: It records four months of statistics on traffic speed, ranging from 3/1/2012 to 6/30/2012, including 207 sensors on the highways of Los Angeles County. The source is available at: https://github.com/liyaguang/DCRNN.
\item LOOP: It is collected from loop detectors deployed on four connected
    freeways (I-5, I-405, I-90 and SR-520) in the Greater Seattle Area. It
    contains traffic state data from 323 sensor stations over the entirely
    of 2015 at 5-minute intervals. The source is available at:
    https://github.com/zhiyongc/Seattle-Loop-Data.

\item Los-loop: This dataset is collected in the highway of Los Angeles
    County in real time by loop detectors. It includes 207 sensors and its
    traffic speed is collected from 3/1/2012 to 3/7/2012. These traffic
    speed data is aggregated every 5 minutes. The source is available at:
    https://github.com/lehaifeng/T-GCN/tree/master/data.

\item TaxiBJ: Trajectory data is the taxicab GPS data and meteorology data
    in Beijing from four time intervals: 1st Jul. 2013 - 30th Otc. 2013,
    1st Mar. 2014 - 30th Jun. 2014, 1st Mar. 2015 - 30th Jun. 2015, 1st
    Nov. 2015 - 10th Apr. 2016. The source is available at:
    https://github.com/lucktroy/DeepST/tree/master/ data/TaxiBJ.

\item SZ-taxi: This is the taxi trajectory of Shenzhen from Jan.1 to
    Jan.31, 2015. It contains 156 major roads of Luohu District as the
    study area. The speed of traffic on each road is calculated every 15
    minutes. The source is available at:
    https://github.com/lehaifeng/T-GCN/tree/master/data.
\item NYC Bike: The bike trajectories are collected from NYC CitiBike
    system. There are about 13000 bikes and 800 stations in total. The
    source is available at: https://www.citibikenyc.com/system-data. A
    processed version is available at:
    https://github.com/lucktroy/DeepST/tree/master/data/ BikeNYC.
\item NYC Taxi: The trajectory data is taxi GPS data for New York City from
    2009 to 2018. The source is available at:
    https://www1.nyc.gov/site/tlc/about/tlc-trip-record-data.page.
\item Q-Traffic dataset: It consists of three sub-datasets: query
    sub-dataset, traffic speed sub-dataset and road network sub-dataset.
    These data are collected in Beijing, China between April 1, 2017 and
    May 31, 2017, from the Baidu Map. The source is available at:
    https://github.com/JingqingZ/BaiduTraffic\#Dataset.
\item Chicago: This is the trajectory of shared bikes in Chicago from 2013 to 2018. The source is available at: https://www.divvybikes.com/system-data.
\item BikeDC: It is taken from the Washington D.C.Bike System. The dataset
    includes data from 472 stations and four time intervals of 2011, 2012,
    2014 and 2016. The source is available at:
    https://www.capitalbikeshare.com/system-data.
\item ENG-HW: It contains traffic flow information from inter-city highways between three cities, recorded by British Government, with a time range of 2006 to 2014. The source is available at: http://tris.highwaysengland.co.uk/detail/trafficflowdata.
\item T-Drive: It consists of tremendous amounts of trajectories of Beijing taxicabs from Feb.1st, 2015 to Jun. 2nd 2015. These trajectories can be used to calculate the traffic flow in each region. The source is available at: https://www.microsoft.com/en-us/research/publication/t-drive-driving-directions-based-on-taxi-trajectories/.
\item I-80: It is collected detailed vehicle trajectory data on eastbound
    I-80 in the San Francisco Bay area in Emeryville, CA, on April 13,
    2005. The dataset is 45 minutes long, and the vehicle trajectory data
    provides the precise location of each vehicle in the study area every
    tenth of a second. The source is available at:
    http://ops.fhwa.dot.gov/trafficanalysistools/ngsim.htm.

\item DiDi chuxing: DiDi gaia data open program provides real and free
    desensitization data resources to the academic community. It mainly
    includes travel time index, travel and trajectory datasets of multiple
    cities. The source is available at: https://gaia.didichuxing.com.

\item Travel Time Index data:

The dataset includes the travel time index of Shenzhen, Suzhou, Jinan, and
Haikou, including travel time index and average driving speed of
city-level, district-level, and road-level, and time range is from 1/1/2018
to 12/31/2018. It also includes the trajectory data of the Didi taxi
platform from 10/1/2018 to 12/1/2018 in the second ring road area of
Chengdu and Xi'an, as well as travel time index and average driving speed
of road-level in the region, and Chengdu and Xi'an city-level. Moreover,
the city-level, district-level, road-level travel time index and average
driving speed of Chengdu and Xi'an from 1/1/2018 to 12/31/2018 is
contained.

\textit{Travel data}:

This dataset contains daily order data from 5/1/2017 to 10/31/2017 in Haikou City, including the latitude and longitude of the start and end of the order, as well as the order attribute of the order type, travel category, and number of passengers.

\textit{Trajectory data}:

This dataset comes from the order driver trajectory data of the Didi taxi
platform in October and November 2016 in the Second Ring Area of Xi'an and
Chengdu. The trajectory point collection interval is 2-4s. The trajectory
points have been processed for road binding, ensuring that the data
corresponds to the actual road information. The driver and order
information were encrypted, desensitized and anonymized.

\end{itemize}

\textbf{Common external data} Traffic prediction is often influenced by a
number of complex factors, which are usually called external data. Here, we
list common external data items.

\begin{itemize}
\item Weather condition: temperature, humidity, wind speed, visibility and
    weather state (sunny/rainy/windy/cloudy etc.)
\item Driver ID:

Due to the different personal conditions of drivers, the prediction will
have a certain impact, therefore, it is necessary to label the driver, and
this information is mainly used for personal prediction.

\item Event: It includes various holidays, traffic control, traffic
    accidents, sports events, concerts and other activities.
\item Time information: day-of-week, time-of-day.

    (1) day-of-week usually includes weekdays and weekends due to the
    distinguished properties.

    (2) time-of-day generally has two division methods, one is to
    empirically examine the distribution with respect to time in the
    training dataset, 24 hours in each day can be intuitively divided into
    3 periods: peak hours, off-peak hours, and sleep hours. The other is to
    manually divide one day into several timeslots, each timeslot
    corresponds to an interval.
\end{itemize}

\section{Experimental analysis and discussions}
\label{sec:experiments} In this section, we conduct experimental studies for
several deep learning based traffic prediction methods, to identify the key
components in each model. To this end, we utilize METR-LA dataset for speed
prediction, evaluate the state-of-the-art approaches with public codes on
this dataset, and investigate the performance limits.

\subsection{Experimental Setup}
In the experiment, we compare the performance of six typical
speed prediction methods with public codes on a public dataset. Table
\ref{method_link} summarizes the links of public source codes for related
comparison methods.
\begin{table}[h]
\centering
\caption{Open source codes of comparison methods.}
\label{method_link}
\begin{tabular}{c|c}
	\hline
	 Approach&Link\\
	
												\hline \hline
STGCN\cite{yu2017spatio}	&\url{https://github.com/VeritasYin/STGCN_IJCAI-18}\\\hline
DCRNN\cite{li2018diffusion}	&\url{https://github.com/liyaguang/DCRNN}\\\hline
ASTGCN\cite{guo2019attention}&\url{https://github.com/guoshnBJTU/ASTGCN-r-pytorch}\\\hline
Graph WaveNet\cite{wu2019graph}&\url{https://github.com/nnzhan/Graph-WaveNet}\\\hline
STSGCN\cite{Song2020Spatial}&\url{https://github.com/Davidham3/STSGCN}\\\hline
GMAN\cite{zheng2019gman}	&\url{https://github.com/zhengchuanpan/GMAN}\\
\hline
	\end{tabular}
	\end{table}

METR-LA dataset: This dataset contains 207 sensors and collects 4 months of
data ranging from Mar 1st 2012 to Jun 30th 2012 for the experiment. 70\% of
data is used for training, 20\% is used for testing while the remaining 10\%
for validation. Traffic speed readings are aggregated into 5 minutes windows,
and Z-Score is applied for normalization. To construct the road network
graph, each traffic sensor is considered as a node, and the adjacency matrix
of the nodes is constructed by road network distance with a thresholded
Gaussian kernel\cite{shuman2013emerging}.

We use the following three metrics to evaluate different models: Mean
Absolute Error (MAE), Rooted Mean Squared Error (RMSE), and Mean Absolute
Percent Error (MAPE).

\begin{equation}\label{15}
  MAE=\frac{1}{\xi}\sum^\xi_{i=1}\left|\hat{y}^i-y^i\right|,
\end{equation}
\begin{equation}\label{14}
  RMSE=\sqrt{\frac{1}{\xi}\sum^\xi_{i=1}\left(\hat{y}^i-y^i\right)^2},
\end{equation}
\begin{equation}\label{16}
  MAPE=\frac{1}{\xi}\sum^\xi_{i=1}\left|\frac{\hat{y}^i-y^i}{y^i}\right|\ast 100\%,
\end{equation}
where $\hat{y}^i$ and $y^i$ denote the predicted value and the ground truth of region $i$ for predicted time step, and $\xi$ is the total number of samples.

For hyperparameter settings in the comparison algorithms, we set their values
according to the experiments in the corresponding literatures (
\cite{yu2017spatio,li2018diffusion,guo2019attention,wu2019graph,zheng2019gman,Song2020Spatial}
).

\begin{table*}
\centering
\caption{Performance of traffic speed prediction on METR-LA.}
\label{speed_performance}
\begin{tabular}{c|c|c|cccccc}
 \hline
\rule{0pt}{12pt} &$T$&Metric&STGCN&DCRNN&ASTGCN&Graph WaveNet&STSGCN&GMAN\\[4pt]
 \hline
 \hline
 \multirow{9}*{\rotatebox{90}{METR-LA}}&\multirow{3}*{15min}& MAE&2.88 &2.77&4.86 &2.69 &3.01 &2.77 \\
                       &                    & RMSE&5.74&5.38&9.27&5.15&6.69 &5.48\\
                       &                    & MAPE&7.62\%&7.30\%&9.21\%&6.90\% &7.27\% &7.25\%\\

                        \cline{2-9}
                       &\multirow{3}*{30min}&MAE &3.47&3.15&5.43&3.07 &3.42 &3.07\\
                       &                    &RMSE&7.24&6.45&10.61&6.22 &7.93 &6.34\\
                       &                    &MAPE&9.57\%&8.80\%&10.13\% &8.37\% &8.49\% &8.35\%\\
                         \cline{2-9}
                       &\multirow{3}*{60min}&MAE &4.59&3.60&6.51&3.53 &4.09 &3.40\\
                       &                    &RMSE&9.40&7.59&12.52&7.37 &9.65 &7.21\\
                       &                    &MAPE&12.70\%&10.50\%&11.64\%&10.01\% &10.44\% &9.72\%\\

\hline
 \end{tabular}
 \end{table*}
\subsection{Experimental Results and Analysis}

In this section, we evaluate the performance of various advanced traffic
speed prediction methods on the graph-structured data, and the prediction
results in the next 15 minute, 30 minute, and 60 minute ($T$=3, 6, 12) are
shown in Table \ref{speed_performance}.

STGCN applied ChebNet graph convolution and 1D convolution to extract spatial
dependencies and temporal correlations. ASTGCN leveraged two attention layers
on the basis of STGCN to capture the dynamic correlations of traffic network
in spatial dimension and temporal dimension, respectively. DCRNN was a
cutting edge deep learning model for prediction, which used diffusion graph
convolutional networks and RNN during training stage to learn the
representations of spatial dependencies and temporal relations. Graph WaveNet
combined graph convolution with dilated casual convolution to capture
spatial-temporal dependencies. STSGCN simultaneously extracted localized
spatio-temporal correlation information based on the adjacency matrix of
localized spatio-temporal graph. GMAN used purely attention structures in
spatial and temporal dimensions to model dynamic spatio-temporal
correlations.

As can be seen from the experimental results in Table
\ref{speed_performance}: First, the attention-based methods (GMAN) perform
better than other GCN-based methods in extracting spatial correlations. When
modeling spatial correlations, GCN uses sum, mean or max functions to
aggregate the features of each node's neighbors, ignoring the relative
importance of different neighbors. On the contrary, the attention mechanism
introduces the idea of weighting to realize adaptive updating of nodes at
different times according to the importance of neighbor information, leading
to better results. Second, the performance of the spectral models (STGCN and
ASTGCN) is generally lower than that of the spatial models (DCRNN, Graph
WaveNet and STSGCN). In addition, the results of most methods are not
significantly different for 15min, but with the increase of the predicted
time length, the performance of the attention-based method (GMAN) is
significantly better than other GCN-based methods. Since most existing
methods predict traffic conditions in an iterative manner, and their
performance may not be greatly affected in short-term predictions because all
historical observations used for prediction are error-free. However, as
long-term prediction has to produce the results conditioned on previous
predictions, resulting in error accumulations and reducing the accuracy of
prediction greatly. Since the attention mechanism can directly perform
multi-step predictions, ground-truth historical observations can be used
regardless of short-term or long-term predictions, without the need to use
error-prone values. Therefore, the above observations suggest possible ways
to improve the prediction accuracy. First, the attention mechanism can
extract the spatial information of road network more effectively. Second, the
spatial-based approaches are generally more efficient than the spectral-based
approaches when working with GCN. Third, the attention mechanism is more
effective to improving the performance of long-term prediction when modeling
temporal correlation. It is worth mentioning that adding an external data
component is also beneficial for performance when external data is available.
\subsection{Computational complexity} To evaluate
the computation complexity, we compare the computation time and the number of
parameters among these models on the METR-LA dataset. All the experiments are
conducted on the Tesla K80 with 12GB memory, the batchsize of each method is
uniformly set to 64, $T$ is set to 12, and we report the average training
time of one epoch. For inference, we compute the time cost on the validation
data. The results are shown in Table \ref{computation}. STGCN adopts fully
convolutional structures so that it is the fastest in training, and DRCNN
uses the recurrent structures, which are very time consuming. Compared to
methods (e.g., STGCN, DCRNN, ASTGCN, STSGCN) that require iterative
calculations to generate 12 predicted results, Graph WaveNet can predict 12
steps ahead of time in one run, thus requiring less time for inference.
STSGCN integrates three graphs at different moments into one graph as the
adjacency matrix, which greatly increases the number of model parameters.
Since GMAN is a pure attention mechanism model that consists of multiple
attention mechanisms, it is necessary to calculate the relation between pairs
of multiple variables, so the number of parameters is also high. Note that,
when calculating the computation time of GMAN, it displays ``out of memory"
on our device, due to the relatively complex design of the model.

\begin{table}[h]
\centering
\caption{Computation cost on METR-LA.}
\label{computation}
\begin{tabular}{l|cc|c}
	\hline
	\multirow{2}{*}{Method} & \multicolumn{2}{c|}{Computation time} & \multirow{2}{*}{Number of parameters} \\
	                      &Training(s/epoch)&Inference(s)& \\
												\hline \hline
STGCN	&49.24&28.13&320143\\
DCRNN	&775.31&56.09&372352\\
ASTGCN	&570.34&25.51&262379\\
Graph WaveNet	&234.82&11.89&309400\\
STSGCN	&560.29&25.56&1921886\\
GMAN   &--&--&900801\\
\hline
	\end{tabular}
	\end{table}

\section{Future Directions}
\label{sec:openproblem} Although traffic prediction has made great progress
in recent years, there are still many open challenges that have not been
fully investigated. These issues need to be addressed in future work. In the
following discussion, we will state some future directions for further
researches.
\begin{itemize}
\item Few shot problem: Most existing solutions are data intensive.
    However, abnormal conditions (extreme weather, temporary traffic
    control, etc) are usually non-recurrent, it is difficult to obtain
    data, which makes the training sample size smaller and learning more
    difficult than that under normal traffic conditions. In addition, due
    to the uneven development level of different cities, many cities have
    the problem of insufficient data. However, sufficient data is usually a
    prerequisite for deep learning methods. One possible solution to this
    problem is to use transfer learning techniques to perform deep
    spatio-temporal prediction tasks across cities. This technology aims to
    effectively transfer knowledge from a data-rich source city to a
    data-scarce target city. Although recent approaches have been proposed
    (\cite{yao2019learning,wei2016transfer,wang2018cross,manibardo2020transfer}
    ), these researches have not been thoroughly studied, such as  how to
    design a high-quality mathematical model to match two regions, or how
    to integrate other available auxiliary data sources, etc., are still
    worth considering and investigating.
\item Knowledge graph fusion: Knowledge graph is an important tool for
    knowledge integration. It is a complex relational network composed of a
    large number of concepts, entities, entity relations and attributes.
    Transportation domain knowledge is hidden in multi-source and massive
    traffic big data. The construction, learning and deep knowledge search
    of large-scale transportation knowledge graph can help to dig deeper
    traffic semantic information and improve the prediction performance.
\item Long-term prediction: Existing traffic prediction methods are mainly
    based on short-to-medium-term prediction, and there are very few
    studies on long-term forecasting (
    \cite{lana2019adaptive,zang2018long,wang2020long}). Long-term
    prediction is more difficult due to the more complex spatio-temporal
    dependencies and more uncertain factors. For long-term prediction,
    historical information may not have as much impact on
    short-to-medium-term prediction methods, and it may need to consider
    additional supplementary information.
\item Multi-source data: Sensors, such as loop detectors or cameras, are
    currently the mainstream devices for collecting traffic data. However,
    due to the expensive installation and maintenance costs of sensors, the
    data is sparse. At the same time, most existing technologies based on
    previous and current traffic conditions are not suited to real-world
    factors, such as traffic accidents. In the big data era, a large amount
    of data has been produced in the field of transportation. When
    predicting traffic conditions, we can consider using several different
    datasets. In fact, these data are highly correlated. For example, to
    improve the performance of traffic flow prediction, we can consider
    information such as road network structure, traffic volume data, points
    of interests (POIs), and populations in a city. Effective fusion of
    multiple data can fill in the missing data and improve the accuracy of
    prediction.
\item Real-time prediction: The purpose of real-time traffic prediction is
    to conduct data processing and traffic condition assessment in a short
    time. However, due to the increase of data, model size and parameters,
    the running time of the algorithm is too long to guarantee the
    requirement of real-time prediction. The scarce real-time prediction
    currently found in the literature \cite{manibardo2020new}, it is a
    great challenge to design an effective lightweight neural network to
    reduce the amount of network computation and improve network speed up.
\item Interpretability of models: Due to the complex structure, large
    amount of parameters, low algorithm transparency, for neural networks,
    it is well known to verify its reliability. Lack of interpretability
    may bring potential problems to traffic prediction. Considering the
    complex data types and representations of traffic data, designing an
    interpretable deep learning model is more challenging than other types
    of data, such as images and text. Although some previous work combined
    the state space model to increase the interpretability of the model
    (\cite{li2019learninginter,pan2019matrix,sen2019think,ziat2017spatio}),
    how to establish a more interpretable deep learning model of traffic
    prediction has not been well studied and is still a problem to be
    solved.
\item Benchmarking traffic prediction: As the field grows, more and more
    models have been proposed, and these models are often presented in a
    similar way. It has been increasingly difficult to gauge the
    effectiveness of new traffic prediction methods and compare models in
    the absence of a standardized benchmark with consistent experimental
    settings and large datasets. In addition, the design of models is
    becoming more and more complex. Although ablation studies have been
    done in most methods, it is still not clear how each component improves
    the algorithm. Therefore, it is of great importance to design a
    reproducible benchmarking framework with a standard common dataset.
\item High dimensionality. At present, traffic prediction still mainly
    stays at the level of a single data source, with less
    consideration of influencing factors. With more collected datasets, we
    can obtain more influencing factors. However, high-dimensional features
    often bring about ``curse of dimensionality" and high computational
    costs. Therefore, how to extract the key factors from the large amount
    of influencing factors is an important issue to be resolved.
\item Prediction under perturbation. In the process of collecting traffic
    data, due to factors such as equipment failures, the collected
    information deviates from the true value. Therefore, the actual sampled
    data is generally subject to noise pollution to varying degrees. The
    use of contaminated data for modeling will affect the prediction
    accuracy of the model. Existing methods usually treat data processing
    and model prediction as two separate tasks. It is of great practical
    significance to design a robust and effective traffic prediction model
    in the case of various noises and errors in the data.
\item The optimal network architecture choice: For a given
traffic prediction task, how to choose a suitable network architecture has
    not been well studied. For example, some works model the historical
    traffic data of each road as a time series and use networks such as RNN
    for prediction; some works model the traffic data of multiple roads as
    2D spatial maps and use networks such as CNN to make predictions. In
    addition, some works model traffic data as a road network graph, so
    network architectures such as GNN are adopted. There is still a lack of
    more in-depth research on how to optimally choose a deep learning
    network architecture to better solve the prediction task studied.

\end{itemize}

\section{Conclusion}
\label{sec:con}

In this paper, we conduct a comprehensive survey of various deep learning
architectures fot traffic prediction. More specifically, we first summarize
the existing traffic prediction methods, and give a taxonomy of them. Then,
we list the representative results in different traffic prediction tasks,
comprehensively provide public available traffic datasets, and conduct a
series of experiments to investigate the performance of existing traffic
prediction methods. Finally, some major challenges and future research
directions are discussed. This paper is suitable for participators to quickly
understand the traffic prediction, so as to find branches they are interested
in. It also provides a good reference and inquiry for researchers in this
field, which can facilitate the relevant research.

\bibliographystyle{IEEEtran}
\bibliography{IEEEabrv,Survey}

\end{document}